\documentclass{llncs}
\usepackage{algorithm2e} 
\usepackage{amssymb} 
\usepackage{graphicx} 
\usepackage{caption}
\usepackage{tikz} 
\usetikzlibrary{arrows}
\usepackage{framed}
\usepackage{comment}
\usepackage{amsmath}
\usepackage{subfigure}
\begin{document}
    \title{Towards Temporal Graph Databases}
    \author{Alexander Campos \and Jorge Mozzino  \and Alejandro Vaisman }
    \institute{Instituto Tecnologico de Buenos Aires\\ \email{jcamposi,jmozzino,avaisman@itba.edu.ar}
 }
    
    \maketitle
    \pagestyle{plain}
    \begin{abstract}
In spite of the extensive literature on graph databases (GDBs), temporal GDBs have not received too much attention so far. Temporal GBDs can capture, for example, the evolution of social networks across time, a relevant topic in data analysis nowadays. In this paper we propose a data model and query language (denoted TEG-QL) for temporal GDBs, based on 
the notion of attribute graphs. This allows a straightforward translation to Neo4J,
a well-known GBD. We present extensive examples of the use of TEG-QL, and comment our
implementation.  
    \end{abstract}
    
    
\thispagestyle{empty}
    
\section{Introduction}
Graphs, and, particularly, attributed graphs~\cite{Wang2014},
are becoming increasingly popular to model different kinds of networks
(e.g., social networks, sensor networks, and the kind) for analysis
in a classical way, and also for performing Online Analytical Processing (OLAP)
on graphs~\cite{Ghrab2015,Wang2014}. Also, these kinds of graphs underlie the 
data model of Neo4J,\footnote{\url{http://www.neo4j.com}}  
probably the most popular graph database for social
network analysis at the time of writing this paper~\cite{Angles2012}, together with AllegroGraph.\footnote{\url{http://www.allegrograph.com}} 
In spite of the fact that social networks are heavily changing 
structures, not much attention has yet been paid to
temporal graph databases  (see Section~\ref{sec:related}). 
In this paper we introduce our approach to this topic, based on attribute graph
data models.  We first present a temporal data model, consisting in a data structure (an attribute graph), and a set of constraints. Then, we sketch a temporal query language, called TEG-QL (for Temporal Graph Query Language), an SQL/SPARQL-like style language, with the idea of facilitating the translation to Cypher, the language that comes with Neo4J. We provide extensive examples of the possible use of TEG-QL. We also describe a prototype we have implemented, allowing time navigation of temporal graphs (and of the result of a TEG-QL query).
 Throughout the paper we will be using a running example consisting in a network
containing two kinds of nodes, representing persons and buildings. Edges in this network are of two kinds:  One, representing friendship relationships between people across time;   the other one  telling the buildings where people had lived through time. Nodes contain information about the  \textsf{name} of the people,  type of building, number of  bedrooms in the apartment, etc. In addition, nodes and edges have a temporal attribute, which is a temporal element   indicating the periods of validity of the node and/or the edge.
 
The rest of the paper is structured as follows:  Section~\ref{sec:related} discusses related work. Then, Section~\ref{sec:datamodel} introduces the data model we propose. Section~\ref{sec:query} presents TEG-QL by example. Section~\ref{sec:imple} briefly explains the basics of our implementation and translation from TEG-QL to Neo4J. We conclude  in Section~\ref{sec:conclu}.

\section{Related Work}
  \label{sec:related}
There is an extensive bibliography on graph database models. This is comprehensively studied in~\cite{Angles2008}. Surprisingly, given the clear need for temporal graph modeling, querying, and analysis, the number of works on the subject is not that large. We survey some work next, and compare against our proposal. 
 
The model we describe in the next section, fits in what are called \textit{Attributed Graphs}, which are appropriate for Online Analytical Processing on graphs~\cite{Wang2014,Ghrab2015}. For example,   attributes in this model allow us to aggregate nodes and edges, provided we define an attribute hierarchy over them. This is one of the reasons of our modeling choice.

A temporal graph model has been presented in~\cite{Catutto13,Catutto13b}, 
where temporal data are organized in so-called frames, namely the finest unit of     temporal aggregation. A frame is associated with a time interval and 
allows to retrieve the status of the social network during such interval. One limitation of that model is that it does not allow registering changes in attributes of the nodes and that frame nodes become too cluttered with edges (one for each actor and for each relationship that existed in that frame). Redundant data are also a problem since each frame is connected to all the existing data, so a frequently  changing graph becomes full of redundant connections.
Opposite to this model, to  keep track of the changes in the value of node attributes, we define \textit{Attribute} and \textit{Value} nodes.
   
Khurana and Deshpande~\cite{Khurana2012,Khurana2013} have studied methods to efficiently query historical graphs. They focus on the particular problem of querying 
the state of a network as of a certain point (snapshot) in time. Also, the authors work on a data model that is based on versioning. Basically, they store the current graph, plus a series of deltas, which contain the graph variation over time. Our model, on the contrary, is based on timestamps, where the complete history is stored in the same graph. This is a typical trade-off problem in temporal databases, discussed also in the temporal XML area~\cite{Rizzolo2008}. 
With   goals different than ours,  Han et al.~\cite{Han2014} presented an engine for temporal graph mining, and  Kostakos~\cite{Kostakos08} show the use of temporal graphs to represent dynamic events.   

\section{Data Model}
\label{sec:datamodel}
We now define the temporal data model supporting our proposal. We base the model in the notion of attribute graphs, that is, graphs whose nodes and edges are annotated with attributes, describing their characteristics. 

\begin{definition}[Temporal graph]
\label{def:tempgraph}
A \textit{temporal graph}  is a structure $G(No, Ne,$ $Na, Nv, E)$ where G is the name of the graph, $E$ is a set of edges, and  $No,$ $Ne,$ $Na,$ and $Nv$ are sets of nodes, denoted   \textit{object nodes}, \textit{edge nodes}, \textit{attribute nodes},and  \textit{value nodes}, respectively.  Every node in the graph is associated with a tuple (\texttt{name},\texttt{interval}). The \texttt{name} represents the content of the node, and the \texttt{interval} represent  the period(s) in which the node is (was) valid  and it is a temporal element. As usual in temporal databases, a special value $Now$ is   used  to represent that the node is currently valid. \qed
\end{definition}

In the definition above, object nodes  represent entities (e.g., \textsf{Person}),  edge nodes  represent  relationships between object nodes (e.g., \textsf{LivesIn}, \textsf{FriendOf},  ), attribute nodes  describe  entities (e.g., \textsf{Name}); Finally, value nodes represent the value of an attribute (e.g., \textsf{Mary}).
The underlying idea is to allow not only to query the graph, but also to perform  OLAP analysis. This is why, instead of placing edges between two object nodes, we define \textit{edge nodes}, which will make aggregation over edges easier. We do not address OLAP analysis of graphs here.

\begin{figure}[t]
            \centering
            \includegraphics[scale=0.6]{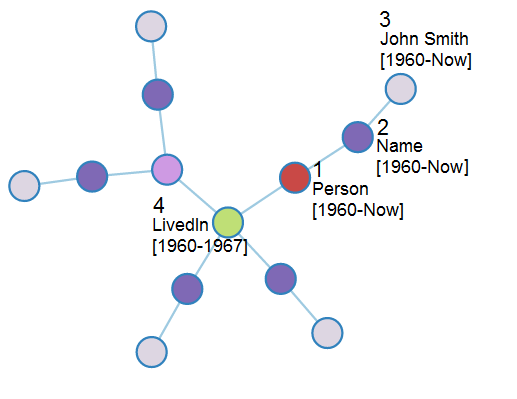}
            \caption{A temporal graph and its different kinds of nodes}
            \label{fig:title_interval}
 \end{figure}

 \begin{example}[Data model]
 Figure~\ref{fig:title_interval} depicts a portion of a graph showing the \texttt{name} and \texttt{interval} properties of the different node types, using our running example.  The properties labeling the nodes are, from top to bottom,  \texttt{id}, \texttt{name} and \texttt{interval}. This way, the node with \texttt{id}=4 (in light green) is an \textit{Edge node}, the  node
with \texttt{id}=1 (in red) is an \textit{Object node}, the node
 with  \texttt{id}=2  (purple) is an \textit{Attribute node}, 
 and the one with \texttt{id}=3 (grey) is a  \textit{Value node}. \qed
 \end{example}
         
Before introducing this graph's constraints, we introduce   some notation. We  denote edge nodes as    $ne\{na, nb\}$, meaning  that $ne$ is an edge node connected to object nodes $na$ and $nb$. An edge will be represented by $e\{na, nb\}$ where $na$ and $nb$ are nodes connected by the edge $e$. An attribute node will be represented as $na\{n\}$ where $n$ is the object or edge node connected to $na$. Finally, we denote a value node as $nv\{na\}$ where $na$ is the attribute node connected to $nv$.
        
\begin{figure}[t]
            \centering
         \subfigure[a]{
         \label{fig:gull} \includegraphics[width=0.35\textwidth]{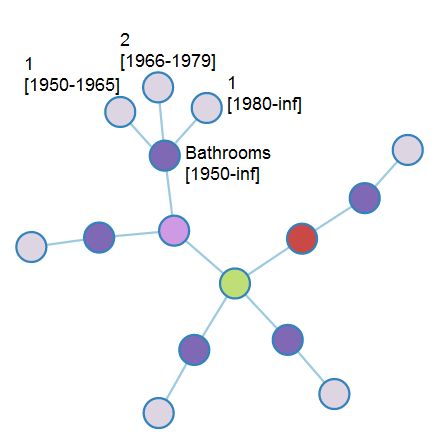}}
            ~
           {
         \label{fig:tiger} \includegraphics[width=0.35\textwidth]{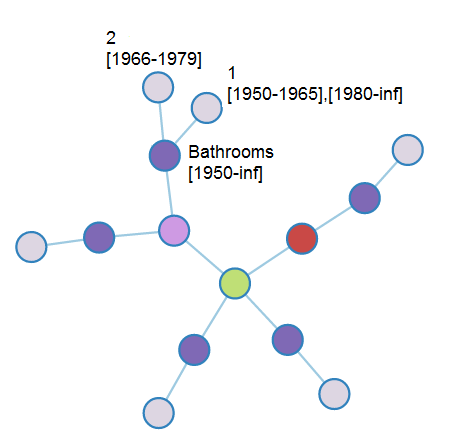}}
\caption{A graph not compliant with constraint \ref{res:same-value} (left); A graph   compliant with constraint \ref{res:same-value} (right)}\label{fig:same-value}
        \end{figure}

\begin{definition}[Constraints]
\label{def:tempconst}
For the graph in Definition \ref{def:tempgraph}, the following constraints hold:
 \begin{enumerate}
 \item $\forall n,n'\in No,~ n= n' \vee n.id \neq n'.id$\label{res:ids-start}
 \item $\forall n, n' \in Ne,~ n= n' \vee n.id \neq n'.id$
 \item $\forall n, n' \in Na,~ n= n' \vee n.id \neq n'.id$
 \item $\forall n, n' \in Nv,~ n= n' \vee n.id \neq n'.id$\label{res:ids-end}
 \item $\forall nv\{na\}, nv'\{na\} \in Nv,~ nv = nv' \vee nv.value \neq nv'.value$ \label{res:same-value}
 \item $\forall n \in No,~ e\{n,n'\} \in E \Rightarrow n' \in Ne \bigcup Na$ \label{res:connection-start}
\item $\forall n \in Ne,~ e\{n,n'\} \in E \Rightarrow n' \in No \bigcup Na$
\item $\forall n \in Na,~ e\{n,n'\} \in E \Rightarrow n' \in No \bigcup Ne \bigcup Nv$ 
\item $\forall n \in Nv,~ e\{n,n'\} \in E \Rightarrow n' \in Nv$ \label{res:connection-end}
 \item $\forall ne \in Ne, ~if~\exists~e\{no, ne\} \land \exists e'\{ne, no'\} 
 \Rightarrow \not\exists e''\in E, \not\exists no''\in No \land no'' \neq no 
 \land no'' \neq no' \land  e''\{no'', ne\} \land e''\{ne, no''\}$   \label{res:cardinality-start}
\item $\forall n \in Na(\exists no \in No \exists e \in E (e(no,n)   
\lor \exists ne \in Ne \land e\{ne, n\}  \land  (
 \not\exists  n' \in (Na \bigcup Ne \bigcup Nv \bigcup No)  \land 
 e' \in E \land e'\{n', n\})$
\item $\forall n \in Nv \land e\{n', n\} \land n \in Na \Rightarrow 
 \not \exists! n'' \in (Na \bigcup Ne \bigcup Nv \bigcup No) \land 
 (e''\{n'', n\} \in E \lor  e''\{n, n''\} \in E$
\item $\exists e\{n,n'\}, e'\{n, n'\} \in E\Rightarrow e = e'$ \label{res:cardinality-end}
\item $\forall ne\{n, n'\} \in Ne, ne.interval \subset n.interval \cap n'.interval$ \label{res:intervals-start}
\item $\forall na\{n\} \in Na, na.interval \subset n.interval$
\item $\forall nv\{na\} \in Nv,  nv.interval \subset nv.interval$
\item $\forall nv\{na\}, nv'\{na\}, nv \neq nv', nv.interval \cap nv'.interval = \emptyset$ \label{res:intervals-end}
        \end{enumerate}
Constraints  \ref{res:ids-start} through \ref{res:ids-end}  state that no two nodes can have the same id. Constraint \ref{res:same-value} requires  coalescing all   nodes with the same value; thus, the interval becomes 
a temporal element which includes all periods where the node had such value.  Figure~\ref{fig:same-value} explains this.  
Constraints \ref{res:connection-start} through \ref{res:connection-end} state how the nodes must be connected, namely: (a) Object nodes can only be connected to edge nodes or attribute nodes; (b) Edge nodes can only be connected to object nodes or attribute nodes; (c) Attribute nodes can be connected to non-attribute nodes; and (d) Value nodes can only be connected to attribute nodes. The cardinalities of these connections is stated by 
   Constraints \ref{res:cardinality-start} through \ref{res:cardinality-end}. Edge nodes must be connected to exactly two different object nodes through exactly one edge, attribute nodes must be connected by only one edge to either an object node or an edge node, and value nodes must only be connected to one attribute node with one edge. Constraint \ref{res:cardinality-end} states that  there cannot be more than one edge between any given pair of nodes.
Constraints \ref{res:intervals-start} to \ref{res:intervals-end} restrict the values of the \texttt{interval} property.
Finally, constraint \ref{res:intervals-end} forces value nodes connected to the same attribute node to have non-overlapping intervals.
 \qed
\end{definition}

\section{TEG-QL: A Query Language for Graphs}
\label{sec:query}   

We now sketch through examples our query language. The syntax of the  language resembles the one of SQL, having the typical \texttt{SELECT-FROM-WHERE} form. 
Queries, as usual in graphs, are based on pattern matching. Thus, the \texttt{FROM} clause contains one or more paths (of fixed or variable length), over which a selection is  performed. The   \texttt{SELECT} clause may either mention just attributes or paths. 
The temporal semantics in embedded in the language, as is normally the case in 
temporal databases~\cite{Tansel93}.  That is, the answer to the query is a temporal graph, although the query may not mention temporal attributes. However, this can be changed by the \texttt{SNAPSHOT} modifier, which allows to retrieve the state of the graph at 
a certain point in time, or the \texttt{IN} modifier, which allows retrieving the status of the graph in a certain interval. There is also the possibility of indicating if we want all the components of the graph, or only the object nodes, and so on. We show  these in the examples below.
    
\begin{figure}[t]
            \centering
          \subfigure{
         \label{fig:select-from-paths} \includegraphics[width=0.17\textwidth]{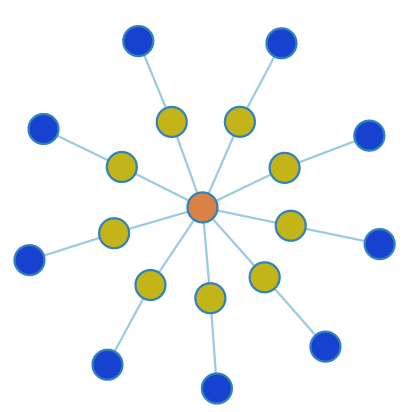}}
            ~~~
            \subfigure{\label{fig:select-with-attributes}
              \includegraphics[scale=0.3]{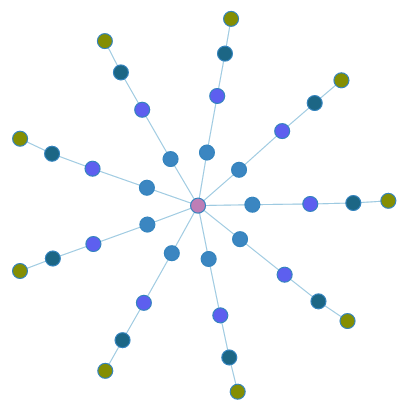}}
            \caption{Query selecting a path (left); Query returning attributes (right)}\label{fig:queries-pat-attributes}
        \end{figure}

We start with a query retrieving object nodes. Consider the query \textit{Buildings where John Smith lives or lived}. This query  just returns the nodes in the graph representing the buildings, and it
 is expressed in TEG-QL  as:

 \begin{verbatim}
           SELECT Building
           FROM Person-LivedIn->Building
           WHERE Person.Name = 'John Smith'
 \end{verbatim}

\vspace*{-0.35cm}
We now move to queries returning paths in the \texttt{SELECT} clause.  
These paths are very similar as the paths in Cypher (the query
language for Neo4J) paths. Consider the query \textit{People and buildings such that a person named John Smith has lived in such buildings.} The predicate we are looking for is \textsf{LivedIn}. The query is expressed in TEG-QL as:

 \begin{verbatim}
            SELECT Person-LivedIn->Building
            FROM Person-LivedIn->Building
            WHERE Person.Name = 'John Smith'
 \end{verbatim}

\vspace*{-0.35cm} 
That is, we take the paths matching the \texttt{FROM} clause, and  filter them using the condition in the WHERE clause. 
Figure \ref{fig:queries-pat-attributes} (left) shows the result.
The center node (in orange) is the \texttt{Person} node that represents John Smith, middle  nodes (yellow) nodes are the edge nodes representing the \textsf{Lived In} relationships; and outer nodes (blue) are the \textsf{Building} nodes.
 
If we just want to return certain attributes, we specify them in a comma-separated list,  between   parenthesis, in the \texttt{SELECT} clause (like in SQL, also a \textsf{*} can be used to represent selecting every attribute of the node). 
The next query shows this: we just want the street of the buildings satisfying the condition in the previous query. This is expressed as follows.

\begin{verbatim}
                SELECT Person-LivedIn->Building(Street)
                FROM Person-LivedIn->Building
                WHERE Person.Name = 'John Smith'
            \end{verbatim}

\vspace*{-0.4cm}
We only return the attribute \textsf{Street} of the nodes of type \textsf{Building}. 
Figure \ref{fig:queries-pat-attributes} (right) shows the result.  From inside out, the nodes are of type  \textsf{Person} (in pink, representing John Smith),  \textsf{LivedIn}, \textsf{Building}, \textsf{Street}, and the value node with the street name (in green). 
      
\begin{figure}[t]
            \begin{center}
              \includegraphics[scale=0.25]{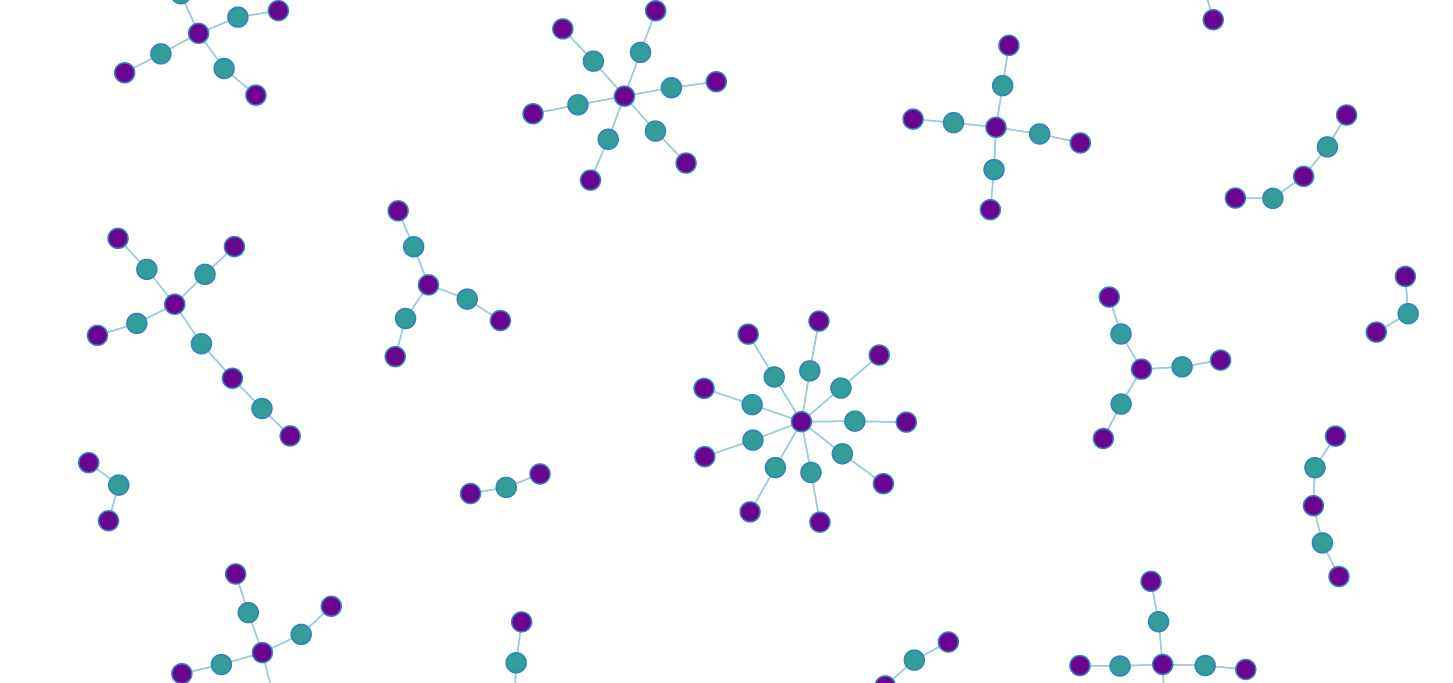}
              \caption{Friends of John Smith}
              \label{fig:john-smith}
            \end{center}
    \end{figure}

A typical query in social network analysis asks for friendship relationships. Let us start with \textit{Friends of someone called John Smith}. Below we show the TEG-QL expression, and Figure~\ref{fig:john-smith} depicts the result, showing  clusters  of people who know someone with the name ``John Smith''. 

             \begin{verbatim}
               SELECT Person-Friend->P2
               FROM Person-Friend->Person as P2
               WHERE Person.Name = 'John Smith'
            \end{verbatim}
            
\vspace*{-0.5cm}    
We mentioned that TEG-QL  supports multiple paths in the \texttt{SELECT} and \texttt{FROM} clauses, allowing also a kind of join between paths. Consider the query \textit{Places where John Smith lived, along with John's friends.} The query reads in TEG-QL:

      \begin{verbatim}
               SELECT * 
               FROM Person-LivedIn->Building,
                    Person as P2-Friend->Person as P3                    
               WHERE Person.Name = 'John Smith' 
                     and P2.Name= 'John Smith'
            \end{verbatim}

\vspace*{-0.5cm}            
This query is equivalent to:  
\vspace*{-0.2cm}
      \begin{verbatim}
               SELECT * 
               FROM Person as P2<-Friend-Person-LivedIn->Building                    
               WHERE Person.Name = 'John Smith' 
            \end{verbatim}
\vspace*{-0.4cm}
Here, the path in the \texttt{FROM} clause reminds   
Cypher syntax. Note that in Neo4J the nodes are of the same type, thus, the path \texttt{Person-Friend-Person} represents, in an abstract way, a traversal of two nodes, while, physically,  the traversal of three nodes.

Variable length paths are also supported, allowing for example, to ask for  the \textit{friends of  John Smith's friends}, also a typical social network query.
\vspace*{-0.2cm}
 \begin{verbatim}
               SELECT * 
               FROM Person-Friend[1..3]->Person                      
               WHERE Person.Name = 'John Smith' 
            \end{verbatim}  

\begin{figure}[t]
            \begin{center}
              \includegraphics[scale=0.3]{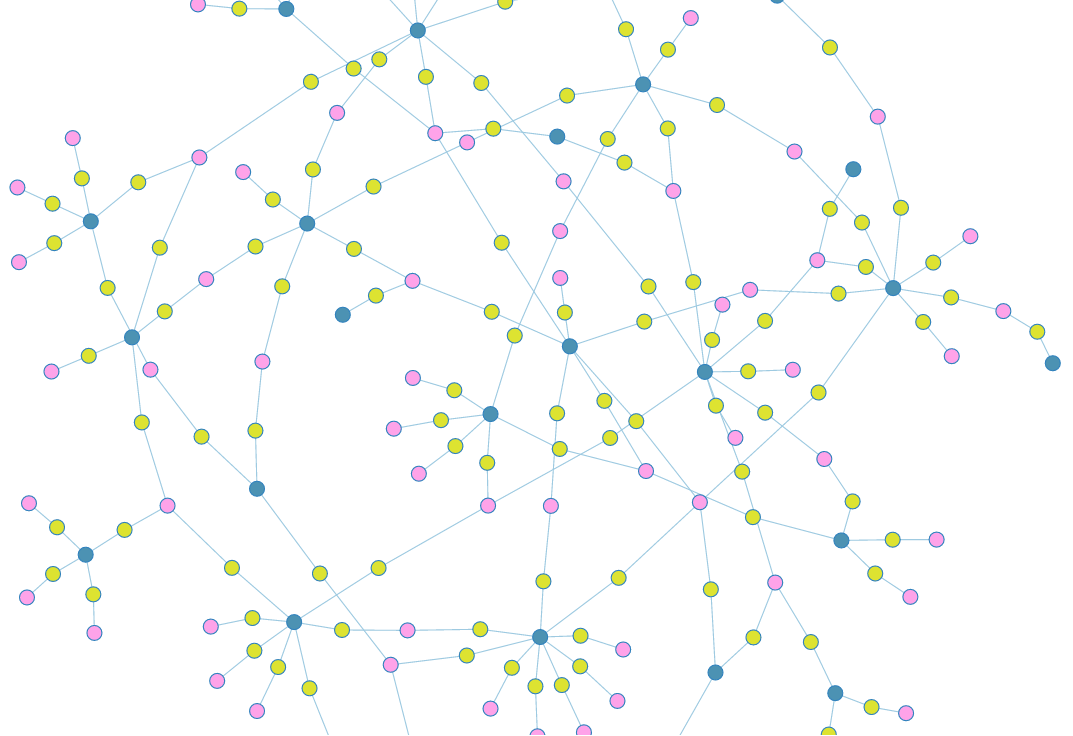}
              \caption{Using the SNAPSHOT clause}
              \label{fig:living_in_1990}
            \end{center}
    \end{figure}

\vspace*{-0.5cm}            
We conclude the section showing the use of the \texttt{SNAPSHOT} and \texttt{IN} modifiers. The query below returns all the people named John Smith, and the buildings where they live, during 1990. 
\vspace*{-0.2cm}
            \begin{verbatim}
                SELECT Person-LivedIn->Building
                FROM Person-LivedIn->Building
                WHERE Person.Name = 'John Smith'
                SNAPSHOT 1990
            \end{verbatim}
\vspace*{-0.5cm}     
Note that we assume a temporal granularity at the \textit{year} level here. We do not get into the details of how to manipulate granularities here. Figure  \ref{fig:living_in_1990} shows the result. Pink nodes represent \textsf{Building} nodes, blue nodes represent \textsf{People} nodes, and yellow nodes represent the \textsf{Lived in} relationship. 

The IN predicate allows selecting an interval where nodes and edges were valid. 
The next query  is similar to the one above, just  selecting those paths existing between 1986 and 1989.
\vspace*{-0.2cm}
 \begin{verbatim}
                SELECT *
                FROM Person-LivedIn->Building
                WHERE Person.Name = 'John Smith'
                IN [1986-1989]
 \end{verbatim}
    
\vspace*{-0.65cm}
\section{Implementation}
\label{sec:imple}    

In this section we explain how TEG-QL queries are translated into Cypher (the Neo4J language), and 
describe the visual interface, that makes use of the temporal model to navigate the graphs across time.  To allow understanding  the process, let us comment on the  Cypher syntax.  Cypher  queries  match paths and then return  parts of that paths. A simple query which returns every node, reads: 
\vspace*{-0.5cm}
 \begin{verbatim}
         MATCH (n)  RETURN n
          \end{verbatim}   
\vspace*{-0.5cm}
Node labels can be specified by modifying the node matcher, like the following query, which returns all the nodes with the \texttt{Object} label.
\vspace*{-0.3cm}
 \begin{verbatim}
         MATCH (n:Object) RETURN n
         \end{verbatim}  
\vspace*{-0.5cm}
In Neo4J, edges are also labeled, so a path can be queried based on these labels.  The following query shows how we match a path of length one between 2 nodes of type person and the relationship representing friendship.   
 \begin{verbatim}
  MATCH (o:Object {title: 'Person'})-->(e:Edge {title: 'Friend'})
       --> (o1:Object:'Person')
  RETURN o,e,o1
    \end{verbatim}
 
\vspace*{-0.8cm}
\subsubsection*{Query Translation}
A TEG-QL query is translated and executed as follows.
First, we translate each path in the \texttt{FROM} clause, checking if each element of each path has an alias, and building the actual Cypher path. An Object node is translated  as \texttt{element.alias:OBJECT title:element.name};   an Edge node, is translated as \texttt{element.name:EDGE title:element.name}. After this, we have our first sentence that,  
for every path in the \texttt{FROM} looks like: 
      \vspace*{-0.2cm}
      \begin{verbatim}
  MATCH (element1.alias:OBJECT {title:element1.name})-->
      (element2.name:EDGE {title:element2.name})-->
      (element3.alias:OBJECT {title:element3.name}) ...
 \end{verbatim}
\vspace*{-0.3cm}
We then   expand the \texttt{SELECT} clause  with the corresponding attributes. If the attributes are empty we do nothing. If the attribute "*" is present, then we want all the attributes. If the attributes are explicit, we translate them one by one. For this, we build a Cypher path from the Object or Edge nodes, adding the Attribute Node and the Value  node(s). For example if we have 
\texttt{Person(Age, Gender)} the translated query will look like this: 
\vspace*{-0.2cm}
     \begin{verbatim}
  MATCH (Person:OBJECT {title:Person})-->(x:ATTRIBUTE {title:Age})
         --> (y:VALUE), 
  MATCH (Person:OBJECT {title:Person})-->(w:ATTRIBUTE {title:Gender})
         --> (z:VALUE)
     \end{verbatim}
\vspace*{-0.3cm}
Finally we address the \texttt{WHERE} statement. We split \texttt{AND}s and \texttt{OR}s and translate each part as follows. If an \textit{id} is present, we produce the path to that id, if it is not already present; if is a constant we translate it as it is. For example if in the \texttt{WHERE} condition we have \texttt{Person.age = 12 AND Person.gender = "Male"}, the translated query will look like this: 
\vspace*{-0.1cm}
\begin{verbatim}
 MATCH (Person:OBJECT {title:Person})--> 
       (x:ATTRIBUTE {title:Age})--> (y:VALUE), 
 MATCH (Person:OBJECT {title:Person})--> 
       (w:ATTRIBUTE {title:Gender})--> (z:VALUE) 
 WHERE y = 12 AND z = "Male" 
 \end{verbatim}
\vspace*{-0.3cm}

Once the query has been executed and the result retrieved, we treat the temporal conditions  \texttt{SNAPSHOT} and \texttt{IN}. This is done by a Java application that interprets the   query and filters the result retrieved by Neo4J. Of course, this is a na\"ive way of treating this part of the query. In future work we will address the problem of query processing. 

\begin{figure}[t]
            \centering
            \subfigure{\label{fig:timeline1}
                \includegraphics[width=0.4\textwidth,height=2.4cm]{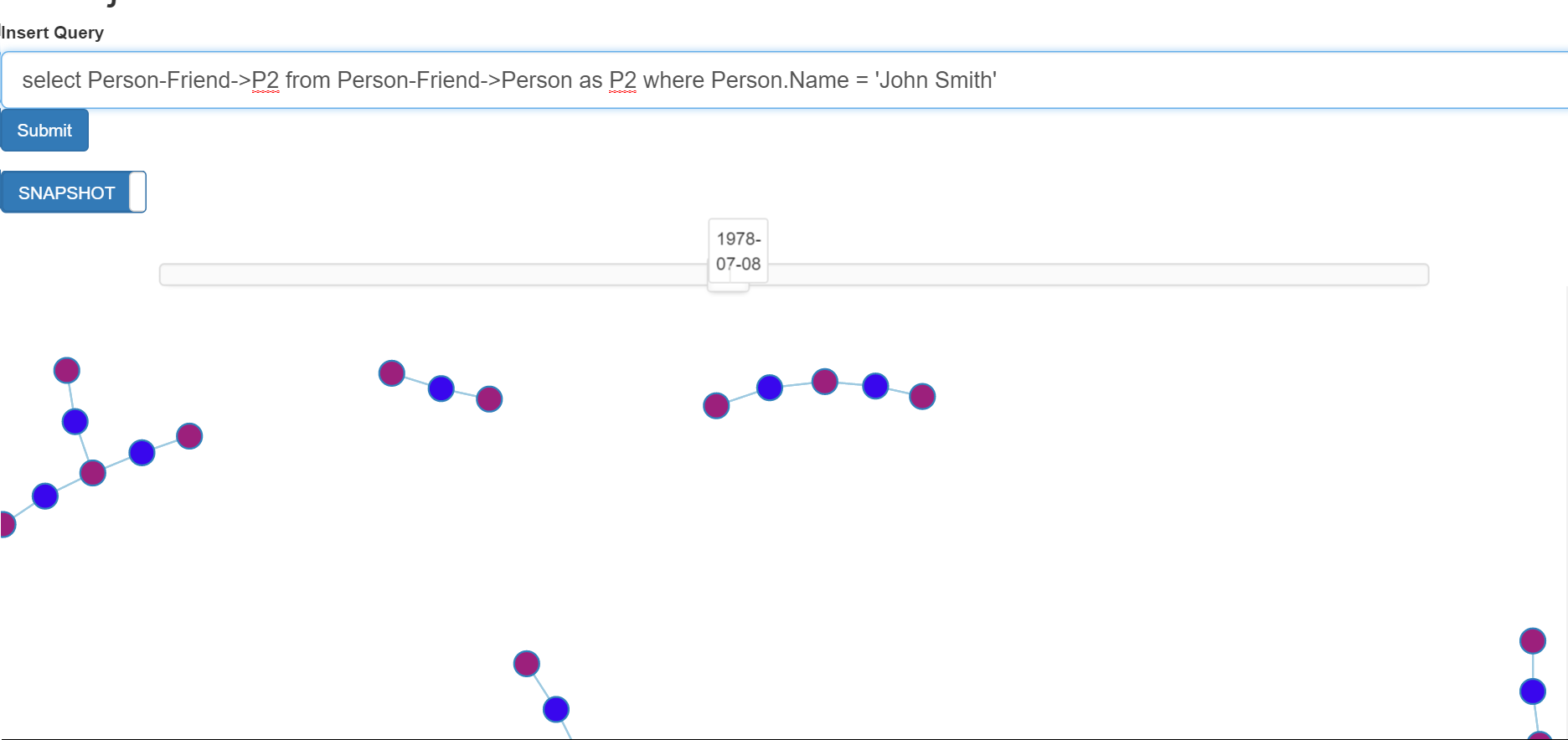}}
            ~
            \subfigure{\label{fig:timeline2}\includegraphics[width=0.4\textwidth,height=2.3cm]{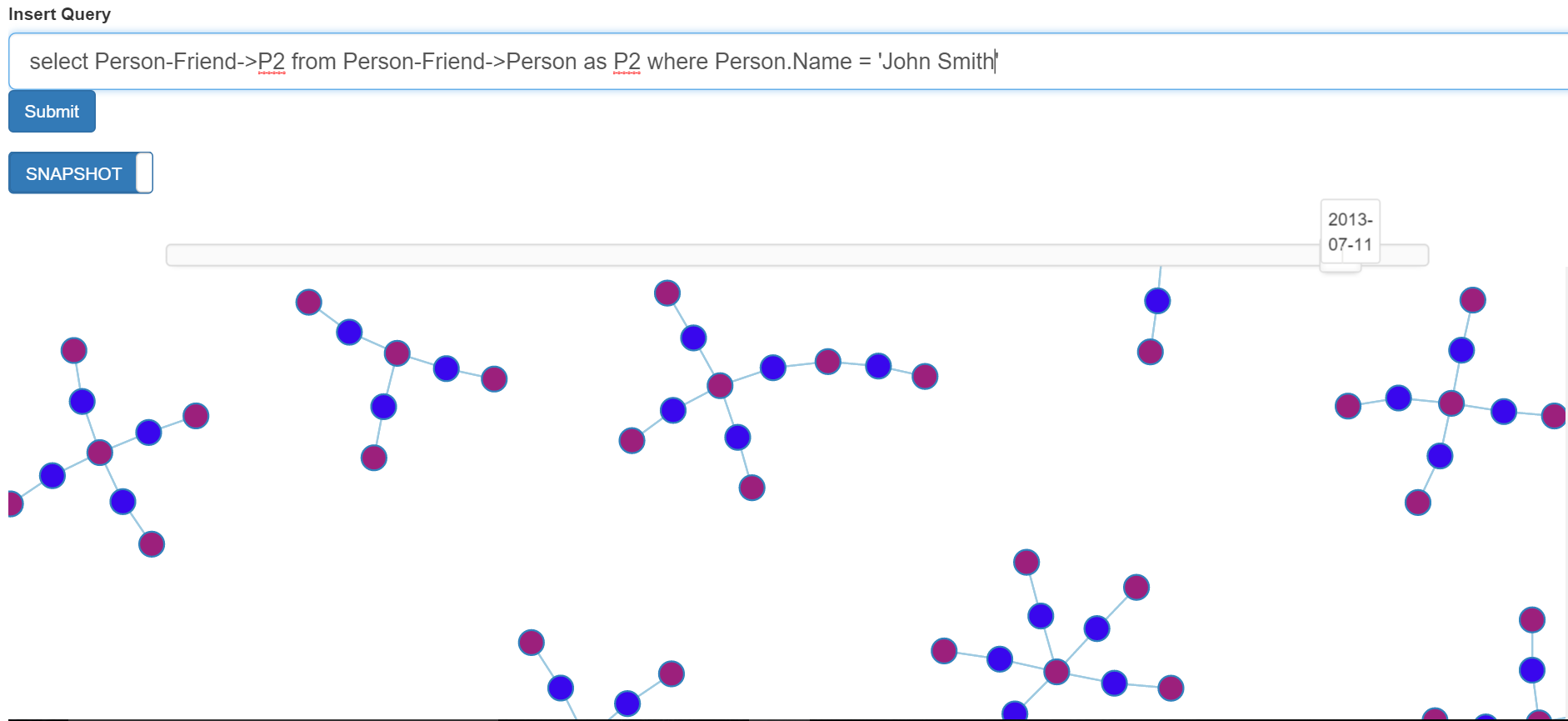}}
            \caption{Visual interface showing two states of the graph}\label{fig:timeline}
        \end{figure}

Finally, we have implemented a prototype to process queries, and a visual interface\footnote{This interface is available at \url{ http://52.37.51.136:8080/}}. This interface   allows the user 
to see the graph (i.e., the result of a query) at a certain point in time (a snapshot), or navigate it across time making use of a sliding bar.
Figure \ref{fig:timeline} depicts two states of a graph , as displayed in the interface, showing the friends of ``our'' John Smith at two points in time.

We ran very preliminary tests over our   na\"{i}ve implementation, using
a T2.Micro on the Amazon cloud, with a  High Frequency Intel Xeon Processors with Turbo up to 3.3GHz CPU, and 1GB RAM. We built a   graph  containing  1000 people nodes, 100 building nodes, 2500 friendship relationships, and   500 lived-in relationships. For the RAM we had available, we could not build larger Neo4J graphs, since Neo4J cannot handle them. Translation times are negligible, as expected; just to give an idea of execution  times, for queries like the ones showed in the example, those times go from 4 to 6 seconds. The reader can try these queries at the
URL mentioned above.

    \section{Future Work}
    \label{sec:conclu}
We have described our approach to model and query temporal graph, which we believe is a relevant problem, for example, in social network analysis,  given the dynamic nature of such networks. Future work will focus on expanding the capabilities of  TEG-QL, and, most of all,  on addressing  the problem of query optimization, for which, we believe that efficient indexing techniques must be developed, opening an interesting research field.

\bibliographystyle{splncs03}

\end{document}